\documentclass{article}
\usepackage{spconf,amsmath,graphicx,cite}
\usepackage{color}
\usepackage{xcolor}
\usepackage{hyperref}
\usepackage{enumerate,url,multirow}
\usepackage{bm}
\usepackage{amsfonts}
\usepackage[normalem]{ulem}
\usepackage{multicol}
\usepackage{algorithm}
\usepackage{algpseudocode}
\def\sH{{\mathsf{H}}}
\def\sT{{\mathsf{T}}}

\newcommand\undermat[2]{
  \makebox[0pt][l]{$\smash{\underbrace{\phantom{%
    \begin{matrix}#2\end{matrix}}}_{\text{$#1$}}}$}#2}

\title{Unsupervised vocal dereverberation \\ with diffusion-based generative models}
\name{\begin{tabular}{@{}c@{}}
Koichi Saito \qquad Naoki Murata \qquad Toshimitsu Uesaka \qquad Chieh-Hsin Lai \\ Yuhta Takida \qquad Takao Fukui \qquad Yuki Mitsufuji
\end{tabular}}

\address{Sony Group Corporation, Tokyo, Japan}
\begin{document}
\ninept

\maketitle
\begin{abstract}

Removing reverb from reverberant music is a necessary technique to clean up audio for downstream music manipulations. 
Reverberation of music contains two categories, \textit{natural reverb}, and \textit{artificial reverb}. 
Artificial reverb has a wider diversity than natural reverb due to its various parameter setups and reverberation types.
However, recent supervised dereverberation methods may fail because they rely on sufficiently diverse and numerous pairs of reverberant observations and retrieved data for training in order to be generalizable to unseen observations during inference.  
To resolve these problems, we propose an unsupervised method that can remove a general kind of artificial reverb for music without requiring pairs of data for training. 
The proposed method is based on diffusion models, where it initializes the unknown reverberation operator with a conventional signal processing technique and simultaneously refines the estimate with the help of diffusion models.
We show through objective and perceptual evaluations that our method outperforms the current leading vocal dereverberation benchmarks.

\end{abstract}
\begin{keywords}
vocal dereverberation, diffusion-based generative models, weighted prediction error
\end{keywords}
\section{Introduction}
\label{sec:intro}
\vspace{-1pt}

Reverb is one of the major audio effects that enable listeners to perceive the spatial characteristics, timbre, and texture of music.
Reverb for music contains not only ``natural reverb'', which has been studied extensively in the context of speech dereverberation problems, but also ``artificial reverb'', which is mostly exploited as an effect for music production~\cite{Koo2021ICASSP, Valimaki2012IEEETASLP}.
In contrast, removing reverbed components from reverberant (wet) music signals is also an important technique for music production. 
Audio engineers and content creators do not merely mix up each signal when creating remixed, remastered, or upmixed music materials from existing contents---they may also apply new kinds of audio effects such as equalization or reverb~\cite{Valimaki2012IEEETASLP,Wilmering2020AS,Zolzer2011book} to original unprocessed (dry) music signals.
However, whenever dry signals are not available, the reverb must be removed from the processed music signals before the desired manipulations can be applied.

To remove reverb from music signals, we need to take both natural and artificial reverb into account. 
Most research in this vein has focused on supervised approaches applied in a data-driven manner that require the preparation of numerous pairs of wet observations and their corresponding dry signals for training.
This tends to get challenging because artificial reverb may have a higher number of variations than natural reverb due to its different parameter setups.
More precisely, the creation of artificial reverb may involve various types of reverberations (e.g., \textit{plate reverb}, \textit{spring reverb}) and  multiple parameters (e.g., \textit{pre delay}, \textit{decay rate}) in a set of reverb having the same RT60. 
Supervised methods generally work well when target samples follow a similar distribution as the training data, but if they deviate from that distribution, the performance may degrade. 
Therefore, to remove artificial reverb, common supervised approaches may not work well for various types of wet signals because the training set may not exhaust the comprehensive data pairs (as discussed in Section~\ref{sec:experiments}).

Indeed, music dereverberation can be formulated as solving a linear inverse problem with a linear degradation operator. Recently, Denoising Diffusion Restoration Model (DDRM)~\cite{Kawar2022NeurIPS}, an unsupervised linear inverse problem solver based on diffusion-based generative models (i.e., diffusion models)~\cite{Ho2020NeurIPS,Song2021ICLR2}, has shown its effectiveness across various image restoration tasks. 
However, DDRM assumes a full knowledge of the linear degradation operator. 
This is problematic as in practical music dereverberation problems, linear degradation operators (i.e., reverberation operators) are generally unknown. 
Hence, directly applying DDRM to the music dereverberation problem may lead to inaccurate retrieval of the dry signal.

To resolve the above problems while (1) avoiding the collection of a large amount of paired data, (2) handling various types of reverb for music production (including artificial reverb), and (3) handling uncertainty of the reverberation operators, we propose an unsupervised method for music dereverberation using diffusion models that is motivated by DDRM. 
Our proposed method contains two key components. 
First, we extend DDRM into a practical scenario with unknown reverberation operators, where weighted prediction error (WPE)~\cite{Nakatani2010IEEETASLP} is used to initially estimate a reverberation operator.
Second, we propose adaptively correcting the initial reverberation operator estimated by WPE to obtain a more accurate one with the help of the predicted dry signal from a diffusion model. 
Our method only needs dry signals for training, which circumvents the needs to prepare the diverse data pairs required by supervised methods. 
We demonstrate that our method effectively dereverbs various types of wet vocal and outperforms unsupervised and supervised benchmarks through comprehensive objective and subjective evaluations on a set of wet vocal test data. 
We refer to \url{https://koichi-saito-sony.github.io/unsupervised-vocal-dereverb/} for examples of the generated audio samples, and reproducing our experiments with the detailed settings of the training configurations.

\section{Related work}
\label{sec:related_works}

\begin{figure*}[tb]
  \centering
  \includegraphics[width=15cm]{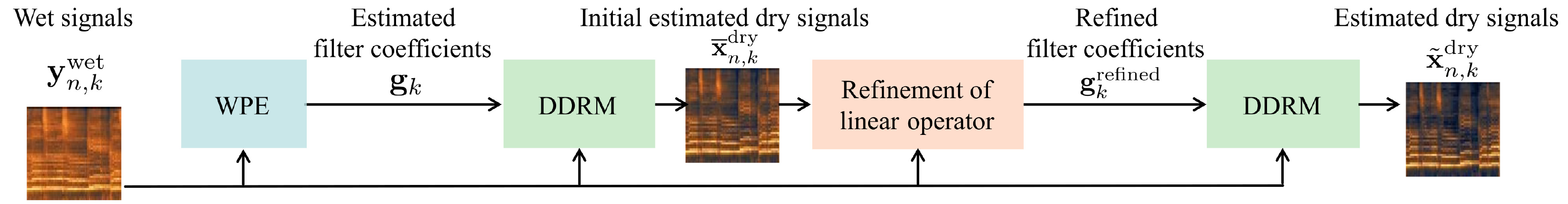}
  \vspace{-6pt}
\caption{Music dereverberation procedure of proposed method.}
\vspace{-10pt}
\label{fig:proposed_method}
\end{figure*}

\vspace{-1pt}
\subsection{Existing research for music dereverberation}
\label{ssec:exisiting_works_for_md}
\vspace{-1pt}

The early unsupervised approaches for music dereverberation focused primarily on signal processing techniques~\cite{Yasuraoka2010ICASSP,Yasuraoka2011ICASSP,Mahkonen2013DAF,Maezawa2014IEEETASLP,Okamoto2012AA,Tsilfidis2011JAES}.
Some of these methods utilized linear prediction~\cite{Mahkonen2013DAF,Maezawa2014IEEETASLP,Okamoto2012AA}, while others were based on the assumption of a source model that directly expresses the harmonic structure of music signals~\cite{Yasuraoka2010ICASSP,Yasuraoka2011ICASSP}.

Some of the recent DNN-based approaches have achieved state-of-the-art performances on vocal and music dereverberation by using an end-to-end DNN~\cite{Koo2021ICASSP} or by combining a conditional generative adversarial network with a diffusion-based vocoder~\cite{Kandpal2022ICASSP}.
Both methods must be trained with pairs of dry and wet signals and are shown to have a decent performance on their test datasets.
However, due to the wide variety of artificial reverb discussed earlier, the effectiveness of these methods may degrade if they are applied to a test dataset that follows a different distribution of wet signals than the one the train dataset was derived from (discussed in more detail in Section~\ref{sec:experiments}).

\vspace{-1pt}
\subsection{Deep generative models for linear inverse problems}
\label{ssec:DDRM}
\vspace{-1pt}

A wide class of image or audio restoration problems can be formulated as linear inverse problems~\cite{ribes2008linear, adler2011audio, arberet2013sparse}. Recent methods~\cite{ulyanov2018deep, shah2018solving, kadkhodaie2021stochastic, kawar2021snips, chung2022improving, chung2022come, Kawar2022NeurIPS} restore the information of data distribution via a deep generative model and use it as a prior to retrieve the signal from an observation.
DDRM~\cite{Kawar2022NeurIPS} aligns with this track in that a highly expressive pre-trained diffusion model is taken as the prior distribution, and DDRM exploits it as a general inverse problem solver with the assumption that the linear data degradation process is known. Since the training of diffusion models requires only clean data (dry signals), not pairs of clean and degraded data, only information about the degradation process is needed during inference.

\vspace{-1pt}
\section{Proposed method}
\label{sec:proposed_method}
\vspace{-1pt}
We propose an unsupervised music dereverberation method for unknown reverberation operators that exploits a pre-trained diffusion model on dry music signals as a prior and does not require pairs of wet and dry signals for training. The proposed method consists of three steps: (1) the initial estimation of the linear operator for the inverse problem via WPE~\cite{Nakatani2010IEEETASLP} (Section~\ref{ssec:estimating_schem_for_reverberant_matix}), (2) the dry music estimation with DDRM (Section~\ref{ssec:solving_linear_inverse_problem}), and (3) the generative model-based refinement of the linear operator (Section~\ref{ssec:h_updating}).


\vspace{-1pt}
\subsection{Problem setting}
\vspace{-1pt}

Let $y_{n, k}^{\text{wet}}\in\mathbb{C}$ be wet signals in the short time Fourier transformation (STFT) domain, where $n$ and $k$ denote the time and frequency indices, respectively. We model the wet signals as
\begin{align}
    y_{n, k}^{\text{wet}} = x_{n, k}^{\text{dry}} + x_{n, k}^{\text{reverb}} + z_{n, k},
    \label{eq:sig_model_music_dereverb}
\end{align}
where $x_{n, k}^{\text{dry}}\in\mathbb{C}$ and $x_{n, k}^{\text{reverb}}\in\mathbb{C}$ are the dry and reverb components included in the wet signal, respectively. Additive noise $z_{n, k}\in\mathbb{C}$ is assumed. The aim of music dereverberation is to estimate the dry signals $x_{n, k}^{\text{dry}}$ from the wet signals $y_{n, k}^{\text{wet}}$.
Here, we assume that a generative model as a prior is available, namely, a pre-trained diffusion model trained on dry signals (see Section~\ref{ssec:solving_linear_inverse_problem}).

\vspace{-1pt}
\subsection{Estimation of linear operator in liner inverse problem}
\label{ssec:estimating_schem_for_reverberant_matix}
\vspace{-1pt}

In this subsection, we revisit WPE~\cite{Nakatani2010IEEETASLP} and introduce its interpretation as a linear inverse problem. 
The idea of WPE is to estimate the reverb components and subtract them from the wet signals. It models the reverb components as the convolution of the filter $\mathbf{g}_{k}=[g_{1, k}, \dots, g_{L, k}]^{\sT}\in\mathbb{C}^{L} $ with the length of $L$ and the wet signals, where $(\cdot)^{\sT}$ denotes the transpose of a matrix (or a vector), as\vspace{-6pt}
\begin{align}
    \hat{x}_{n, k}^{\text{dry}} = y_{n, k}^{\text{wet}} - \sum_{l=1}^{L}g_{l, k}^{*}y_{n-D-l+1, k}^{\text{wet}},
    \label{eq:WPEfiltering}
\end{align}
where $\hat{x}_{n, k}^{\text{dry}}$ is the estimate of dry signals, $(\cdot)^{*}$ denotes the complex conjugate, and $D$ is the prediction delay. The filter is obtained on the basis of various assumptions regarding dry signals, e.g., dry signals at time indices $n$ and $n'$ are assumed to be mutually uncorrelated when $|n-n'|>\delta$ for a certain constant $\delta > 0$.  For more details, refer to the original paper~\cite{Nakatani2010IEEETASLP}. This filter $\mathbf{g}_{k}$ will be refined adaptively with the help of a diffusion model (see Section~\ref{ssec:h_updating}). With the obtained filter, the estimate of the dry signals is obtained by rewriting Eq.~\eqref{eq:WPEfiltering} as follows:\vspace{-6pt}
\begin{align} 
    \hat{\mathbf{x}}_{n, k}^{\text{dry}} = \left(\tilde{\mathbf{I}}-\mathbf{G}_{k}\right)\mathbf{y}_{n, k}^{\text{wet}}, 
    \label{eq:WPE2Linear}
\end{align}
where $\mathbf{y}_{n, k}^{\text{wet}}=[y_{n, k}^{\text{wet}}, \dots, y_{n-D-L-m+1, k}^{\text{wet}}]^{\sT}$, \\$\hat{\mathbf{x}}_{n, k}^{\text{dry}}=[\hat{x}_{n, k}^{\text{dry}}, \dots, \hat{x}_{n-m, k}^{\text{dry}}]^{\sT}$, and $m$ is the number of processed samples. $\tilde{\mathbf{I}}$ and $\mathbf{G}_{k}$ are Toeplitz matrices defined as 
\begin{align}
    \tilde{\mathbf{I}} &= \begin{bmatrix} \mathbf{I}_{m\times m}, \mathbf{0}_{m\times(D+L-1)}
    \end{bmatrix},\\
    \mathbf{G}_{k} &= \begin{bmatrix}
                            & g_{1, k}^{*} & g_{2, k}^{*} & \cdots       & g_{L, k}^{*} & 0            & \cdots & 0 \\ \vspace{-2pt}
    \mathbf{0}_{m\times D}  & 0            & g_{1, k}^{*} & g_{2, k}^{*} & \cdots       & g_{L, k}^{*} & \ddots & \vdots \\\vspace{-2pt}
                            & \vdots       & \ddots       & \ddots       & \ddots       &              & \ddots & 0 \\\vspace{-2pt}
    \hspace{-12pt} \undermat{D}{           &} \undermat{m-1}{0            & \cdots       & 0            &} \undermat{L}{g_{1, k}^{*} & g_{2, k}^{*} & \cdots & g_{L, k}^{*}}
    \end{bmatrix},
\end{align}
where $\mathbf{I}_{m'\times m'}$ denotes the identity matrix of size $m'$, and $\mathbf{0}_{m' \times n'}$ is the $m'\times n'$ rectangular matrix with all the elements $0$. 

Now, the estimated dry signals $\hat{\mathbf{x}}_{n, k}^{\text{dry}}$ can be interpreted as the least-squares solution of the following linear inverse problem:
\begin{align}
        \mathbf{y}_{n, k}^{\text{wet}}&=\left(\tilde{\mathbf{I}}-\mathbf{G}_{k}\right)^{\dagger}\hat{\mathbf{x}}_{n, k}^{\text{dry}} + \mathbf{z}_{n, k},
\end{align}
where $\mathbf{z}_{n, k}=[z_{n, k},\dots,z_{n-D-L-m+1, k}]^{\sT}$. $\mathbf{A}^{\dagger}$ is the pseudo-inverse of a matrix $\mathbf{A}$ and is defined as $\mathbf{A}^{\dagger} = \mathbf{A}^{\sH}\left(\mathbf{A}\mathbf{A}^{\sH}\right)^{-1}$, where $(\cdot)^{\sH}$ denotes the Hermitian transpose of the matrix. 

\vspace{-1pt}
\subsection{DDRM for solving linear inverse problems}
\label{ssec:solving_linear_inverse_problem}
\vspace{-1pt}

\begin{algorithm}
\caption{Proposed algorithm for music dereverberation}\label{alg:proposal}
\begin{algorithmic}
\Require Wet signal $\mathbf{y}_{n, k}^{\text{wet}}$, \\ Pre-trained diffusion model $p_{\theta}(\mathbf{x}^{\text{dry}})$, Number of refinements of the linear operator $N_{\text{refine}}$, Step size $\alpha$, and Regularization parameter $\lambda$
\Ensure Estimated dry signal $\tilde{\mathbf{x}}_{n, k}^{\text{dry}}$
\State Estimate the filter coefficient $\mathbf{g}_{k}$ with WPE~\cite{Nakatani2010IEEETASLP}
    \State Estimate the dry signal $\overline{\mathbf{x}}_{n, k}^{\text{dry}}$ with DDRM~\cite[Eq.~(8)]{Kawar2022NeurIPS}
\For {$i=1$ to $N_{\text{refine}}$}
    \State Refine $\mathbf{g}_{k}$ with~Eq.\eqref{eq:update_lo}
\EndFor
\State Estimate the dry signal $\tilde{\mathbf{x}}_{n, k}^{\text{dry}}$ with DDRM on the refined $\mathbf{g}_{k}^{\text{refined}}$
\end{algorithmic}
\end{algorithm}

Our approach for estimating dry signals is motivated by DDRM~\cite{Kawar2022NeurIPS}.
It solves linear inverse problems in an unsupervised way that requires the linear operator only during inference and hence, pairs of the wet and dry signals are not needed during training for supervision. 
Therefore, music dereverberation with artificial reverb may benefit from this mechanism when the reverberation operation is known.

We review how the prior diffusion models are obtained. First, we train a diffusion model on the dry signal dataset in the STFT domain. We denote a set of $x_{n, k}^{\text{dry}}$ as $\mathbf{x}^{\text{dry}}$. Diffusion models~\cite{Ho2020NeurIPS,Song2021ICLR2,Lai2022Arxiv} are generative models with a Markov chain $\mathbf{x}_{T}^{\text{dry}}\rightarrow \cdots\rightarrow\mathbf{x}_{t}^{\text{dry}}\rightarrow\cdots\rightarrow\mathbf{x}_{0}^{\text{dry}}=\mathbf{x}^{\text{dry}}$ represented by the following joint distribution:
\begin{align}
    p_{\theta}(\mathbf{x}_{0:T}^{\text{dry}}) = p_{\theta}^{(T)}(\mathbf{x}_{T}^{\text{dry}})\prod_{t=0}^{T-1}p_{\theta}^{(t)}(\mathbf{x}_{t}^{\text{dry}}|\mathbf{x}_{t+1}^{\text{dry}}),
\end{align}
where only $\mathbf{x}_{0}^{\text{dry}}$ is used to generate samples for $\mathbf{x}^{\text{dry}}$. For the training, a fixed approximated posterior is introduced to evaluate an evidence lower bound (ELBO) on the maximum likelihood objective:
\begin{align}
    q(\mathbf{x}_{1:T}^{\text{dry}}|\mathbf{x}_{0}^{\text{dry}}) = q^{(T)}(\mathbf{x}_{T}^{\text{dry}}|\mathbf{x}_{0}^{\text{dry}})\prod_{t=0}^{T-1}q^{(t)}(\mathbf{x}_{t}^{\text{dry}}|\mathbf{x}_{t+1}^{\text{dry}}, \mathbf{x}_{0}^{\text{dry}}),
\end{align}
and we adopt the following ELBO objective, which is induced from the Gaussian parameterization for $p_{\theta}$ and $q$: 
\begin{align}
    \mathbb{E}_{t, q(\mathbf{x}_0^{\text{dry}}, \mathbf{x}_{t}^{\text{dry}})}\left[ \gamma_{t}\|\mathbf{x}_{0}^{\text{dry}}-f_{\theta}^{(t)}(\mathbf{x}_t^{\text{dry}}) \|_{2}^{2}\right],
\end{align}
where $f_{\theta}^{(t)}$ is a neural network that characterizes $p_{\theta}$ and estimates a noiseless data from a noisy data $\mathbf{x}_{t}^{\text{dry}}$, and $\gamma_{t}$ are positive weighting coefficients determined by $q$.

The inference of DDRM with a pre-trained diffusion model is based on \cite[Eq.~(8)]{Kawar2022NeurIPS}. In particular, DDRM requires the singular value decomposition (SVD) of the linear operator, as $(\tilde{\mathbf{I}}-\mathbf{G}_{k})^{\dagger} = \mathbf{U}_{k}\mathbf{\Sigma}_{k}\mathbf{V}_{k}^{\sH}$. In our case, we only require the SVD of $(\tilde{\mathbf{I}}-\mathbf{G}_{k})$, since the SVD of the pseudo-inverse becomes $\mathbf{V}\mathbf{\Sigma}^{\dagger}\mathbf{U}^{\sH}$ if the SVD of the original matrix is $\mathbf{U}\mathbf{\Sigma}\mathbf{V}^{\sH}$.
Using the pre-trained diffusion-based model $p_{\theta}(\mathbf{x}^{\text{dry}})$ and the SVD of the linear operator, DDRM generates samples $\overline{\mathbf{x}}_{n, k}^{\text{dry}}$~\cite[Eq.~(8)]{Kawar2022NeurIPS} that are consistent with the linear inverse problem in Eq.~\eqref{eq:WPE2Linear}.

\vspace{-1pt}
\subsection{Generative model-based refinement of linear operator}
\label{ssec:h_updating}
\vspace{-1pt}

Since WPE estimates the filter coefficients $\mathbf{g}_{k}$ on the basis of relatively simple assumptions, e.g., the correlation property of dry signals, it does not necessarily provide reasonable filter coefficients, which are required in DDRM for the linear operator. 
We therefore refine the filter coefficients utilizing the dry signal $\overline{\mathbf{x}}_{n, k}^{\text{dry}}$ estimated by DDRM.
After obtaining the estimates, we further refine the filter coefficients as\vspace{-3pt}
\begin{align}
        \mathbf{g}_{k} \leftarrow \mathbf{g}_{k} - \alpha \nabla_{\mathbf{g}_{k}}\left( \|\overline{\mathbf{x}}_{n, k}^{\text{dry}}- (\tilde{\mathbf{I}}-\mathbf{G}_{k})\mathbf{y}_{n, k}^{\text{wet}}\|_2^{2}+\lambda\|\mathbf{g}_{k}\|_{2}^{2}\right),
    \label{eq:update_lo}
\end{align}
where $\alpha$ is a step size and $\lambda$ is a regularization parameter. Note that $\mathbf{G}_{k}$ is parameterized by $\mathbf{g}_{k}$ and the derivative with respect to $\mathbf{g}_{k}$ is tractable. With the refined $\mathbf{g}_{k}^{\text{refined}}$, the DDRM procedure is executed again to generate higher-quality results $\tilde{\mathbf{x}}_{n, k}^{\text{dry}}$. 
The proposed dereverberation algorithm is summarized in Fig.~\ref{fig:proposed_method} and Algorithm~\ref{alg:proposal}.

\vspace{-1pt}
\section{Experiments}
\label{sec:experiments}

\vspace{-1pt}
\subsection{Dataset}
\label{ssec:dataset}
\vspace{-1pt}


To examine the effectiveness of the proposed method, we conducted both quantitative and subjective evaluations on wet vocal signals. 
The pre-trained diffusion model was trained with only dry vocal signals from the NHSS dataset~\cite{Sharma2020Arxiv}, which contains $100$ English pop songs ($20$ unique songs) of ten different male and female singers. 
The total signal duration is $285.24$ minutes. 

As a test dataset, we prepared $3600$ wet vocal signals, (ten hours in total) by adding artificial reverb to dry vocal signals from another dataset called the NUS-48E corpus~\cite{Duan2013APSIPA}.
This corpus contains $48$ English pop songs ($20$ unique songs) of different male and female singers. 
The total signal duration is $169$ minutes. 
Each song for both training and testing is sampled at $44.1$~kHz and features monaural recording.
As artificial reverb, we used all the presets for vocal in the FabFilter Pro-R plug-in~\footnote{\url{https://www.fabfilter.com/products/pro-r-reverb-plug-in}}.
There are $19$ kinds of vocal reverb presets in total.
We prepared wet vocal signals by first making $48 \times 19$ wet vocal signals and dividing them into $10$-second samples, and then randomly selecting $3600$ signals from among them.

\vspace{-1pt}
\subsection{Experimental settings}
\label{ssec:experimental_settings}
\vspace{-1pt}

The implementation of our method and the network architecture of the pre-trained diffusion model were mostly based on the code provided by the authors of the DDRM paper~\footnote{\url{https://github.com/bahjat-kawar/ddrm}}.
We slightly modified some parts as follows. 
We converted each audio input into a complex-valued STFT representation using a window size of $1024$, a hop size of $256$, and a Hann window.
Further, we cut the direct current component of the input signals and input them as $2$-channeled $512 \times 512$ image data to follow the original input configurations. The first channel corresponds to the real value and the second to the imaginary value.
We modified the original U-Net~\cite{Ronneberger2015ICM} architecture of the pre-trained model used on DDRM by adding a time-distributed fully connected layer~\cite{Choi2020ISMIR} to the last layer of every residual block.
For the training, we reduced the size of the diffusion model to have the fewer trainable parameters ($31.3$~M), and the training took less than three days using one NVIDIA A100 GPU.

For the inference, the parameters of WPE, DDRM, and the proposed refinement were set as follows. 
For WPE, we set $L=150$ and $D=8$, with the number of iterations set to one.
For DDRM, we followed the same notations defined as in \cite[Eq.~(8)]{Kawar2022NeurIPS} and set $\eta=0.7, \eta_{b} = 0.2$, and $\sigma_{y} = 1.0 \times 10^{-6}$, with the number of sampling steps set to $20$. 
For the proposed refinement, we set $\alpha = 1.0 \times 10^{-6}, \lambda=1.0$, and $N_{\text{refine}} = 10000$.


In addition, to explore the limitation of DDRM and our methods, we tested cases where a reverberation operator was able to be approximated from an oracle dry signal of test data $\mathbf{x}_{n, k}^{\text{test dry}}$ for a given wet observation of test data $\mathbf{y}_{n, k}^{\text{test wet}}$. We obtained this operator by minimizing $\|\mathbf{y}_{n, k}^{\text{test wet}} - \mathbf{H}_{k}\mathbf{x}_{n, k}^{\text{test dry}}\|_2^{2}$ with respect to $\mathbf{H}_{k}$ over the reverberation operator.
\vspace{-1pt}
\subsection{Baselines}
\label{ssec:baselines}
\vspace{-1pt}

We evaluated the proposed method against three baselines.

\noindent \textbf{Reverb conversion (RC)}: A state-of-the-art end-to-end DNN-based method for vocal dereverberation. We used the original code and the pre-trained model\footnote{The original code and the pre-trained model were shared by Junghyun Koo from the Artificial Intelligence Institute at Seoul National University. Mr. Koo also assisted with the discussion of the RC results of our experiment.}, which was trained with the pairs of $44.1$~kHz wet and dry vocal signals. 
Note that the wet signals were reverbed with artificial reverb taken from the different professional reverb plug-ins from those of our test dataset~\cite{Koo2021ICASSP}.
We input pairs of wet and dry signals since this method needs them for dereverberation.

\noindent \textbf{Music enhancement (ME)}: A supervised method to denoise and dereverb music signals based on diffusion models~\cite{Kandpal2022ICASSP}. 
We used both the original code and the pre-trained model specified in the paper. 
Since ME was trained with pairs of $16$~kHz reverberant noisy and clean music signals containing vocal signals, we evaluated this method at $16$~kHz. 

\noindent \textbf{WPE}: An unsupervised method for speech dereverberation~\cite{Nakatani2010IEEETASLP}. 
We set $L=200, D=8$, and the number of iterations to three.

\vspace{-1pt}
\subsection{Quantitative evaluation}
\label{ssec:quantitative_evaluation}

\begin{table}[t]
    \caption{Results of quantitative evaluation. $\ell_{1}$ loss means $\ell_{1}$ loss of magnitude spectrogram in STFT domain. Proposed and Proposed+ denote our methods without and with proposed refinement. DDRM w/ est-Oracle denotes case where DDRM was given an approximated oracle reverberation operator.}
        \label{tab:objective_resutls}
        \centering
\resizebox{8.5cm}{!}{
    \begin{tabular}{c|c|cccc}
    \hline
     Methods & Manner & $\ell_{1}$ loss $\downarrow$ & FAD $\downarrow$ \\
     \hline
     Wet (Unprocessed) & -- & $0.114$ & $13.7$ \\ \hline 
     RC~\cite{Koo2021ICASSP} & Supervised & $0.117$ & $13.9$  \\
     ME~\cite{Kandpal2022ICASSP} & Supervised & $0.484$ & $14.7$ \\ \hline 
     WPE~\cite{Nakatani2010IEEETASLP} & Unsupervised & $0.103$ & $10.1$ \\
     \textbf{Proposed} & Unsupervised & $0.102$ & $9.85$ \\
     \textbf{Proposed+} & Unsupervised & \bm{$0.100$} & \bm{$9.69$} \\\hline
      DDRM w/ est-Oracle & -- & $0.079$ & $4.44$ \\ \hline
    \end{tabular}
}
\end{table}

We evaluated the dereverberation performance of the proposed method without the refinement of an initial reverberation operator (Proposed) and with it (Proposed+) by computing two objective metrics. 
The first was the $\ell_{1}$ loss of the amplitude spectrogram in the STFT domain between a dereverbed and dry signal.
The other metric was the Fréchet audio distance (FAD)~\cite{Kilgour2018Arxiv} between the set of dry and dereverbed signals. 
Since the VGGish~\cite{S_Hershey2017ICASSP}, which is the pre-trained classification model of FAD, is originally trained with the $16$~kHz audio samples, we dowmsampled all the signals to $16$~kHz and computed FAD.

Table~\ref{tab:objective_resutls} lists the scores of each measurement.
Both versions of the proposed method showed better scores than all the baselines on both metrics, and Proposed+ scored better than Proposed.
These results demonstrate the effectiveness of our method and that utilizing the proposed refinement leads to a better dereverberation performance.
The large gap between the scores of Proposed and DDRM w/ est-Oracle indicates that the estimation accuracy of reverberation operators significantly affects the dereverberation performance of our method.
Thus, a better estimation of the initial reverberation operators and a better refinement of them will lead to a better dereverberation performance, which we leave to future work.

RC and ME did not perform well at all, which may according to that the distribution of their training dataset did not cover that of test dataset.
Indeed, the training wet signals of ME and RC were created using only simulated natural reverb with some background noise~\cite{Kandpal2022ICASSP} and different artificial reverb plug-ins from those of our test dataset~\cite{Koo2021ICASSP}, respectively.
Another reason RC did not work well is that this model may not have been trained with enough pairs of wet and dry signals. 
Since RC was originally meant for not only dereverberation but also converting styles of reverb, it was also trained with the pairs of two different wet signals (see~\cite[Section $4.1$]{Koo2021ICASSP}).

\subsection{Listening test}
\label{ssec:listening_test}
We also conducted a listening test using the multiple stimuli with hidden reference and anchor (MUSHRA) method~\cite{MUSHRA}.
A total of $18$ participants took part in the test on the webMUSHRA platform~\cite{Schoeffler2018webMUSHRA}. 
The participants were presented with ten kinds of signals (one for the practicing part and nine for the test part) selected randomly from the test dataset in Section~\ref{ssec:quantitative_evaluation} and trimmed to six seconds each.
The results of the practicing part were removed from the aggregation of the results.
Each web page contained a wet vocal signal as a reference and participants were asked to rate five different signals according to how much they felt the reverberant components were removed compared to the reference. 
The five signals were composed of the outputs of RC, WPE, Proposed+, a reference wet signal as a hidden reference, and a dry vocal signal as a hidden anchor. 
As with the results in Section~\ref{ssec:quantitative_evaluation}, the outputs of ME and Proposed were not included here considering the burden on participants.

\begin{figure}[tb]
  \centering
  \centerline{\includegraphics[width=7.5cm]{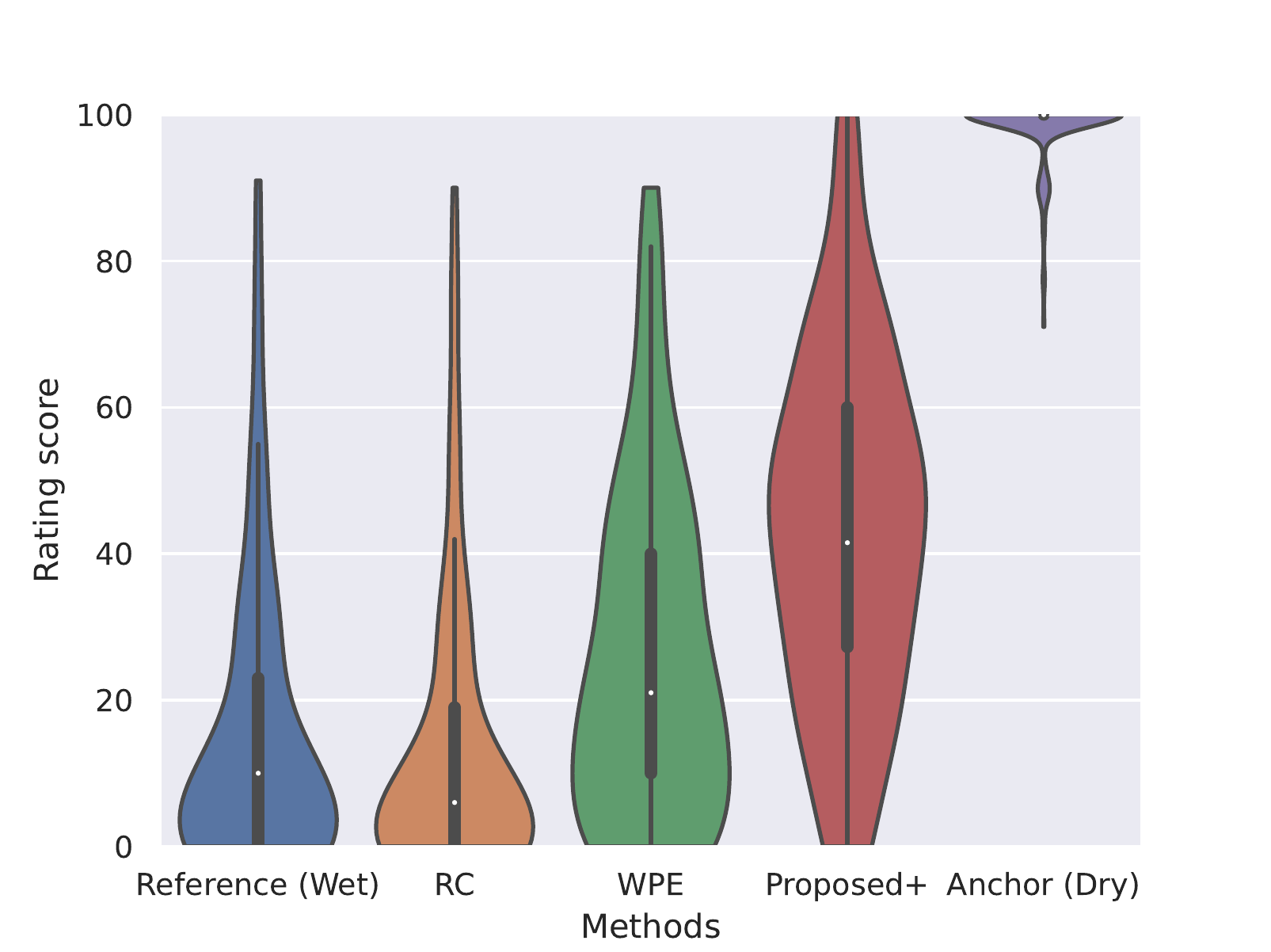}}
\caption{Violin plots of listening test scores. White dots denote median of scores and top and bottom of vertical bold lines denote first and third quartiles, respectively.}
\label{fig:listening_test}
\end{figure}

Figure.~\ref{fig:listening_test} shows violin plots of the listening test results.
Dry signals showed the highest score, which confirms that participants were able to judge which were wet and which were dry, as expected.
Proposed+ showed the best score, outperforming both RC and WPE, which demonstrates the effectiveness of the proposed method, in terms of the perceptual metric.

\section{CONCLUSION}
\label{sec:conclusion}

In this work, we proposed an unsupervised method that can remove general reverb for music including artificial reverb without requiring diverse pairs of data for training.
We extended DDRM to a practical use case with unknown reverberation operators. First, we initialize the reverberation operator with an estimation from WPE.
Second, we adaptively refined the initial reverberation operator estimated by WPE to get a more accurate one with the help of a dry signal predicted by a diffusion model. 
The results of both objective and perceptual evaluations demonstrate that our method outperforms the current leading vocal dereverberation benchmarks.
\vfill\pagebreak
\bibliographystyle{IEEEbib}
\fontsize{8.9pt}{9.0pt}
\selectfont{}

\vfill\pagebreak

 \onecolumn
\section{Appendix}
\label{sec:appendix}
In this section, we additionally demonstrate the detailed settings of the training configurations of the pre-trained diffusion model. 
The network architecture of the pre-trained diffusion model is mostly based on the code provided by the authors of the DDRM paper~\footnote{\url{https://github.com/bahjat-kawar/ddrm}}. Especially, we explain our modifications about the input representation and the network architecture in Section~\ref{ssec:experimental_settings}.
The pre-trained diffusion models in the DDRM code are from this \textit{guided-diffusion} GitHub repository~\footnote{\url{https://github.com/openai/guided-diffusion}}~\cite{Dhariwal2021NeurIPS}.
The hyperparameters for the training of the diffusion model are in Table~\ref{tab:appendix}.
We also incorporate an adaptive group normalization~\cite{Dhariwal2021NeurIPS} into each residual block.
We train the model using AdamW~\cite{Loshchilov2019ICLR} with $\beta_{1}=0.9$ and $\beta_{2}=0.999$ in $16$-bit precision~\cite{Micikevicius2018ICLR}. We use an exponential moving average over model parameters with a rate of $0.9999$~\cite{Song2020NeurIPS}.
     
\begin{table}[h]
    \caption{Hyperparameters for training diffusion model. We followed same notations defined in~\cite[Table 11]{Dhariwal2021NeurIPS}}
        \label{tab:appendix}
        \centering
\resizebox{7.0cm}{!}{
    \begin{tabular}{lc}
    \hline 
     Diffusion steps & $4000$ \\
     Noise schedule & cosine~\cite{Nichol2021Arxiv} \\
     Model size & $31.3$~M \\  
     Channels & $64$  \\
     Depth & $2$ \\  
     Channels multiple & $1, 1, 2, 2, 4, 4$ \\
     Heads & $2$ \\
     Attention resolution & $32, 16$ \\
     BigGAN up/downsample & \checkmark \\
     Dropout & $0.0$ \\
     Batch size & $6$ \\
     Iterations & $370$K\\
     Learning rate & $1.0\times 10^{-4}$ \\
     \hline
    \end{tabular}
}
\end{table}

\end{document}